\documentclass[prc,twocolumn]{revtex4}
\usepackage{amsmath,amssymb,epsfig,bm,mathrsfs}
\newcommand{\be}{\begin{equation}}
\newcommand{\ee}{\end{equation}}
\newcommand{\bea}{\begin{eqnarray}}
\newcommand{\eea}{\end{eqnarray}}

\newcommand{\ie}{{\it i.e.}}
\newcommand{\eg}{{\it e.g.}}

\begin{document}
\title{
Thermal evolution of massive compact objects with dense quark cores
}

\author{Daniel Hess, Armen Sedrakian}
\affiliation{Institute for Theoretical Physics,
J.-W. Goethe University, D-60438  Frankfurt-Main, Germany
}

\date{\today}
 
\begin{abstract}
We examine the thermal evolution of a sequence of compact objects
containing low-mass hadronic  and high-mass quark-hadronic stars
constructed from a microscopically motivated equation of state. The 
dependence of the cooling tracks in the temperature versus age 
plane is studied on the variations of the gaplessness parameter  (the 
ratio of the pairing gap for red-green quarks to the electron chemical 
potential) and the magnitude of blue quark gap. The pairing 
in the red-green channel is modeled assuming an inhomogeneous
superconducting phase to  avoid tachionic instabilities and
anomalies in the specific heat; the blue colored condensate is
modeled as a Bardeen-Cooper-Schrieffer (BCS)-type color superconductor. 
 We find that massive stars containing 
quark matter cool faster in the neutrino-cooling era if one of the
colors (blue) is unpaired and/or the remaining colors (red-green)
are paired in a inhomogeneous gapless superconducting state. 
The cooling curves show significant variations along the sequence,  
as the mass (or the central density) of the models is varied. This 
feature provides a handle for fine-tuning the models to fit the data 
on the surface temperatures of same-age neutron stars. In the
late-time photon cooling era we observe inversion in the temperature 
arrangement of models, \ie, stars experiencing fast neutrino cooling 
are asymptotically hotter than their slowly cooling counterparts. 
\end{abstract}

\maketitle

\section{Introduction}
\label{sec:1}
After the initial phase of rapid (of order of days to weeks)
cooling from temperatures $T \sim 50$ MeV down to 0.1 MeV,  a neutron
star settles in a thermal quasiequilibrium state which
evolves  slowly over the time scales $10^3-10^5$ yr down to temperatures
$T \sim 0.01$ MeV.  The cooling rate of the star during this period is
determined  by the processes of neutrino emission from dense matter,
whereby the neutrinos, once produced, leave the star without further
interactions. The understanding of the cooling processes that take place
during  this neutrino radiation era is crucial for the interpretation
of the data  on surface temperatures of neutron stars. While the long
term  features of the thermal evolution of neutron stars are
insensitive  to the initial rapid cooling stage ($t\le $ years), the
subsequent route in the  temperature versus time plane, 
which includes in addition to the neutrino emission era the 
late-time ($t \ge 10^5$ yr)  photoemission era, strongly depends on
the emissivity of matter during the neutrino-cooling era~\cite{Glen:1980,Page:2005fq,Yakovlev:2007vs,Sedrakian:2006mq}.

The neutrino emission during the neutrino-cooling era chiefly
originates from the core of a compact object, \ie,  from matter at
and above densities  $\sim 0.16$ fm$^{-3}$ and is sensitive to its 
composition.
Hadronic  matter emits neutrinos mainly via the beta and inverse beta
decay processes and the charge neutral processes of Cooper pair breaking. 
The former are suppressed by pairing correlations (\ie, the emergence of a
gap in the quasiparticle spectrum separating the ground state from the
excited states below certain critical temperature), whereas the latter
are the consequence of the formation of Cooper pair condensates~\cite{Sedrakian:2006mq}.

The densities in the centers of massive compact objects can be by an order 
magnitude greater than the nuclear saturation density. The quark
substructure of the baryons will play an increasingly important role
as larger densities are reached in progressively massive objects. The
recent timing observations of binary millisecond pulsar J1614-2230
imply that compact objects with masses of order of 2 solar masses
exist in nature~\cite{Demorset:2010}.

It is the purpose of this work to examine the cooling behavior of
massive compact objects containing cores of deconfined interacting
quark matter. Specifically, we shall assume that above certain density
the quarks are in continuum states filling a Fermi sphere. The
mechanism by which the neutrinos are produced in quark matter are
basically the same as in the baryonic matter, \ie, beta decays of $d$
quarks and inverse beta decays via electron capture on a $u$
quark. Because of our assumption of filled Fermi spheres and
attractive interaction among quarks in dominant color antisymmetric
states, the pairing of quarks - known as color superconductivity -
needs to be taken into account \cite{Rajagopal:2000wf,Alford:2001dt,
  Hong:2000ck,Rischke:2003mt,Shovkovy:2004me,Alford:2007xm,Wang:2009xf}.
In this paper, we shall concentrate on simple $ud$-quark condensates
and explore the possibility that these condensates are in one of the
gapless phases with broken spatial symmetry~\cite{Alford:2000ze,
  Bowers:2002xr,Muther:2002ej,Kiriyama:2006ui,
  Casalbuoni:2005zp,He:2006vr,Mannarelli:2008zz,Sedrakian:2009kb,Huang:2010bf}.
The homogeneous gapless color superconductors are not treated because
they develop tachionic instabilities in the ground
state~\cite{Huang:2004bg} and their specific heat has anomalous
temperature dependence~\cite{Sedrakian:2006mt}.

Thermal components in the photon spectra from a handful of pulsars
have been identified. The temperatures inferred from these measurements
point toward  a dichotomy in the evolution of isolated neutron stars
- some of them are clearly much hotter than their same-age 
counterparts. The scatter in the data over the temperature-age plane could be
correlated with certain static (mass, radius) and/or dynamic (spin,
$B$-field) parameter(s) of pulsars. Such correlations are not easy to
resolve theoretically, as there could be a number of factors
influencing the emitted spectrum (\eg,  the properties of the emitting
regions could be sensitive to the variation of the magnetic fields at
the surface of the star and its composition). 

One of the aims of this work is to understand the correlation between
the mass and the surface temperature of a cooling compact object for a
class of models which support high-mass ($M\sim 2 M_{\odot}$)
stars. Specifically, we shall study below the models constructed in
Ref.~\cite{Ippolito:2007hn}. The sequences of static and rotating
configurations of Ref.~\cite{Ippolito:2007hn} have maximal masses of
the order $2M_{\odot}$ with radii of the order of 12 km. The internal
structure of these models is described in Ref.~\cite{Knippel:2009st}, 
where it is shown that their massive members contain quark cores with
radii of the order of 7~km and masses $\le 1 M_{\odot}$.

Previous work studied the cooling of quark-hadronic stars with
different input physics and varying degree of realism.  From
simulations of cooling of compact stars with a mixed phase of quarks
and hadrons, in which the neutrino emission processes are uniformly
suppressed by a gap in quark spectrum, one may conclude that quark
matter is invisible if the gap in quark matter is large and
indistinguishable from hadronic matter if the gap is
small~\cite{Page:2000wt}. In simulations of the cooling compact
objects of 1.4 $M_{\odot}$ with quark cores in the color-flavor-locked
(CFL) phase, the cooling is too fast to accommodate the observational
data~\cite{Blaschke:2000dy}.  Because all the fermions in the CFL
phase are gapped, the quark core remains hot for longer periods of
time (up to 100 yr) compared to the phases containing ungapped
fermions in the quark phase.  The cooling simulations of
Ref.~\cite{Alford:2004zr} of 1.4~$M_{\odot}$ toy models having core
quark matter in the gapless version of the CFL phase show that the
cores of such stars would be much hotter than their nuclear
counterparts in the photon cooling era. This is because such matter has 
larger specific heat due to quadratic rather than linear dependence of the
quasiparticle spectrum on the momentum.  The thermal behavior of
Larkin-Ovchinnikov-Fulde-Ferrell quark matter was studied for 1.4
$M_{\odot}$ toy models, and it was concluded that such phases will lead
to rapid cooling of compact stars with rates that are comparable to the
rate found for the unpaired quark matter~\cite{Anglani:2006br}. The 
quark stars that contain two-flavor quark matter in a homogeneous
state, possibly in a mixed state with normal quark matter, would cool
too fast to account for the data~\cite{Grigorian:2004jq}. The
possibility that the normal quarks additionally pair in some channel
fits the data~\cite{Grigorian:2004jq,Blaschke:2006gd} if the masses of
the models are varied in a range $1.0-1.75 ~M_{\odot}$ and some
suitable density dependence is assigned to the smaller gap.

In this work, we are concerned with the thermal evolution of a sequence
of compact objects, parametrized by their central density or mass,
which contains large-mass ($M\le 2 M_{\odot}$) members with quark
cores. Our models have realistic backgrounds, being derived from a
microscopically motivated equation of state and the general
relativistic structure equations. One aim of this study is to quantify
the changes in the cooling behavior of stars along the sequence,
focusing mainly on the new features that arise with the onset of quark
matter in the cores of high-mass models. We assume quark matter
of light $u$ and $d$ quarks in beta equilibrium with electrons; the
$s$ and other quark flavors are suppressed by their large masses.  The
pairing among the $u$ and $d$ quarks occurs in two channels: the
red-green quarks are paired in a condensate with gaps of the order of
the electron chemical potential; the blue quarks are paired with (smaller)
gaps of order of keV, which is comparable to core temperature during
the neutrino-cooling
epoch~\cite{Alford:2002rz,Buballa:2002wy,Schmitt:2004et}.  The
neutrino emission in the case of single-flavor, but multicolor pairing
is as in normal quark matter~\cite{Schmitt:2005wg} and differs from
the single-color case studied here.

 A homogenous red-green condensate, when in the gapless BCS regime,
 develops tachionic instabilities, which manifest themselves in
 negative Meissner masses~\cite{Huang:2004bg}.  Furthermore, the
 thermodynamic quantities of a homogeneous gapless superconductor are
 generally anomalous at low temperatures; in particular, for large
 enough asymmetries the gap disappears at a lower critical temperature
 and the specific heat experiences a
 jump~\cite{He:2006vr,Sedrakian:2006mt}.  Recent studies show that
 a realization of inhomogeneous superconductors, the Fulde-Ferrell
 phase, is preferred in color superconducting quark matter to the
 homogenous BCS phase at sufficiently low temperature and weak
 coupling for arbitrary flavor asymmetries~\cite{Sedrakian:2009kb} as
 well as when color and charge neutrality are
 enforced~\cite{Huang:2010bf}.  Furthermore, the anomalies in the
 thermodynamic quantities are cured when the inhomogeneous phases are
 treated (this conjecture is rigorously proven for the so-called
 Fulde-Ferrell phase~\cite{He:2006vr,Sedrakian:2009kb,Huang:2010bf}).

 Motivated by these observations and to avoid anomalies related to the
 homogeneous phases, we shall assume a generic inhomogeneous state and
 model the neutrino emission and specific heat in a phenomenological
 manner (see Sec.~\ref{sec:2}).  For the red-green condensate, we shall
 use a parameterization of neutrino emissivity in terms of the
 gaplessness parameter, as proposed in Ref.~\cite{Jaikumar:2005hy}.
 Thus, the second aim of this work is to understand the systematics of
 the cooling tracks in the temperature versus age plane when the
 ``gaplessness'' parameter, \ie, the ratio of the pairing gap for
 red-green quarks to the electron chemical potential is varied.  The
 magnitude of the gap in the spectrum of blue quarks is another
 parameter that will be varied. We do not attempt to fit the data by
 fine-tuning the available parameters (the mass of the star, the
 gaplessness parameter, and the gap for blue quarks); rather, we would
 like to find general trends that could provide orientation once the
 observational data demands such fits.

This article is organized as follows. In Sec.~\ref{sec:2}, we describe
the models of purely hadronic and quark-hadronic stars, the pairing
patterns in quark matter, and the microscopic input needed 
for studying their thermal evolution. Section~\ref{sec:3} studies the
thermal evolution of these models. Here, we discuss the key
approximations used to evolve the models in time and discuss 
the variations in cooling behavior as the central density of the
models as well as the pairing parameters are changed. Our conclusions
are collected in Sec.~\ref{sec:4}. Some preliminary results were reported
in Refs.~\cite{Knippel_Diplom,Sedrakian:2009uu,Sedrakian:2010bd}.

\section{Quark cores and neutrino emission processes}
\label{sec:2}

\subsection{Equation of state, composition, and pairing}
We start with a brief description of the models underlying our study. 
The low-density nuclear matter equation of state is based on a
relativistic model of nuclear matter in a homogeneous core and 
an inhomogeneous low-density phase constituting the crust of the star
(see Ref.~\cite{Ippolito:2007hn} for details).
The high-density matter equation of state is derived from the effective
Nambu-Jona-Lasinio four-fermion interaction
model which incorporates the pairing interactions among quarks.
The two equations were matched by a   Maxwell construction, \ie, the
requirement that the pressures of the two phases at the point of
transition to the deconfined quark phase should match at a fixed
baryo-chemical potential. The high-density quark phase pairs the up ($u$)
and down ($d$) quarks. We will assume that  strange quarks 
(because of their large mass) are not important at the densities
relevant to our models. With this equation of state as an input, the models
of low-density purely hadronic and high-density quark-hadronic stars
were constructed by solving the general relativistic structure
equations. The masses and radii of various components are shown
for a set of models in Table~\ref{table:1}. It is assumed that the
transition from the hadronic core to the crust occurs at $0.5\rho_0$,
where $\rho_0= 2.8 \times 10^{14}$ g cm$^{-3}$ is the nuclear
saturation density. In the isothermal-interior approximation described
below, the temperature gradients are concentrated in the envelope of the
star;  the transition from isothermal interior to the nonisothermal envelope 
occurs at the density $\rho_m = 10^{10}$ g cm$^{-3}$ and the radius
$R_{cr}$. Table~\ref{table:1} lists the values of the inner radius
$R_{cr}$ of the models.

\begin{figure}[tbh] 
\begin{center}
\includegraphics[width=8cm,height=7cm]{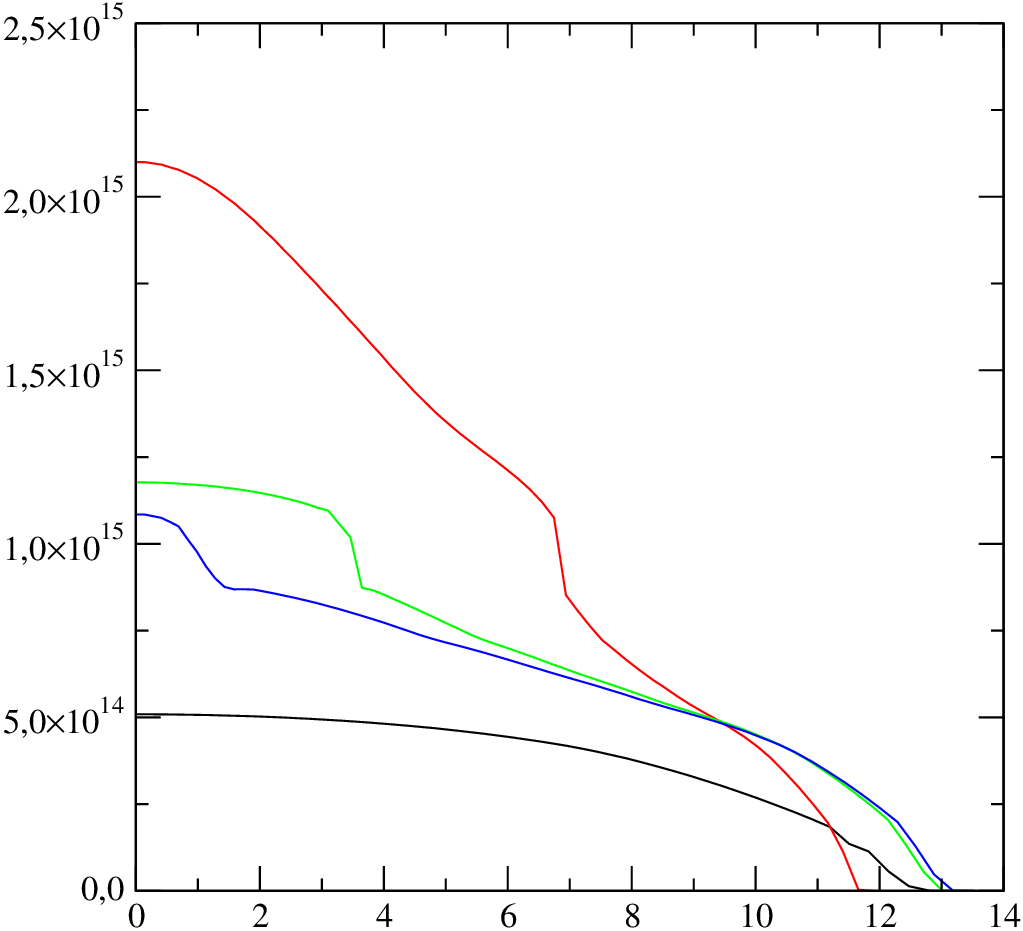}
\caption{ Dependence of the density of the stellar matter on the
  radial coordinate for static, spherically symmetrical stars. The
  central densities and radii of the stars can be read off from the
  respective axes.  The observed jumps in the density profiles of
  high-mass stars are a consequence of the onset of the quark 
  matter in the core.
}
\label{fig:1}
\end{center}
\end{figure}

Figure~\ref{fig:1} shows the density profiles of four models selected
from the set above, whose thermal evolution will be studied in
Sec.~\ref{sec:3}.  One of these is a purely hadronic model with central
density $\rho_{c,14} = 5.1$, where $\rho_{c,14} = \rho_c /( 10^{14}$ g
cm$^{-3})$. The remaining three models contain a quark matter core and
have central densities $\rho_{c,14} = 10.8,\,11.8,\,21.0$.  The model
with $\rho_{c,14}= 21.0$ has the maximum mass for the sequence of the
stars defined by the equation of state above, \ie, for larger central
densities the stable branch of equilibrium, nonrotating,
configurations terminates.
The quark-hadron transition is seen in the density jump at the radius
at which the phase transition occurs. (Note that the pressure and
the baryonic chemical potential are continuous at the point of
transition by construction.) The chosen models provide a good coverage 
of the possibilities provided by the underlying equation of state. In particular,
a substantial fraction of the mass and volume of the model with largest
central density is occupied by quark matter. 
\begin{table}[ht]
  \caption{Masses and the radii of models constructed from the
    equation of state adopted in this work.  The low-density models 
    are purely hadronic, whereas the high-density models contain 
    quark cores. The central densities of the models are given in
    units of $10^{14}$ g cm$^{-3}$, the radii in kilometers, and the 
    masses in units of the solar mass. The columns from left to right
    are: the central density, the
    radius of the quark core,  the quark-plus-hadronic core,  
    the isothermal interior and the star, the masses of the quark 
    core, the hadronic core, 
    the crust, and  the star.}    
		\begin{tabular}{ccccccccc} 
\\
\hline
\hline\\
	   $\rho_{c,14}$ & $R_{Q}$ & $R_{Q+H}$ & $R_{cr}$ & $R$
           &$M_{Q}$  & $M_{H}$ & $M_{cr}$ & $M$ \\
\\
\hline
\hline\\
	    3.6&-&9.48&13.20&13.80&-&0.43&0.11&0.54\\
	    5.1&-&11.47&13.39&13.53&-&1.03&0.07&1.10\\
	    8.2&-&12.57&13.55&13.57&-&1.81&0.04&1.85\\
\\
\hline
\\
	    10.8&0.68&12.54&13.49&13.5&0.001&1.81&0.040&1.91\\	    
	    11.8&3.41&12.40&13.31&13.32&0.09&1.80&0.04&1.93\\
	    15.0&5.84&11.86&12.54&12.55&0.52&1.46&0.03&2.01\\
	    21.0&6.77&11.34&11.91&11.92&0.89&1.14&0.02&2.05\\
\\
	  \hline\hline\\
		\end{tabular}
	\label{table:1}  
\end{table}

The combination of charge neutrality and beta equilibrium implies the
presence in the stellar matter of electrons, whose chemical potential
obeys the condition $ \mu_e = \mu_d -\mu_u.  $ Because of the shift in
the chemical potentials of the quarks of different flavors, the pairing
pattern differs from the conventional Bardeen-Cooper-Schrieffer (BCS)
superconductors by the fact that the paired fermions are drawn
from different Fermi surfaces. Under these conditions the pairing is
``gapless.'' \ie, there exist segments on the Fermi surface where the
gap separating the excited states of the system from the ground state
disappears. The gapless phases could be either homogeneous (in the
sense that the spatial $O(3)$ symmetry is intact) or inhomogeneous, in
which case the symmetry is broken down to a certain symmetry subgroup.

\subsection{Neutrino emissivities and other input}

Quark matter consisting of two light flavors cools via the beta decay  (Urca) reactions 
\be 
d\to u+e+\bar \nu,\quad u+e\to d + \nu,
\ee
where $\nu$ and $\bar \nu$ stand for neutrino and antineutrino. For
unpaired quarks and to leading order in the strong coupling constant
$\alpha_s$ the emissivity of the process {\it per quark color} is given by~\cite{Iwamoto:1980eb}
\be\label{eq:iwamoto}
\epsilon_{\beta} =\frac{914}{945}\tilde G^2 p_d p_up_e\alpha_sT^6 ,
\ee
where $\tilde G = G\cos \theta$ is the weak coupling constant,
$\theta$ is the Cabibbo angle, $p_d$, $p_u$, and $p_e$ are the Fermi momenta of
down quarks, up quarks, and electrons. Non-Fermi-liquid corrections to
the Urca rate above~\cite{Schafer:2004jp} were neglected in our study.

The quark pairing modifies the temperature dependence of the process
(\ref{eq:iwamoto}). In the BCS-type superconductors, the process is
suppressed nearly linearly for $T\simeq T_c$ and exponentially for
$T\ll T_c$, where $T_c$ is the critical
temperature~\cite{Sedrakian:2004qd}.  In gapless
superconductors the emergence of the new scale $\delta\mu =
(\mu_d-\mu_u)/2$ leads to two essentially different possibilities~\cite{Jaikumar:2005hy}, which
are distinguished by the value the dimensionless parameter $\zeta =
\Delta/\delta\mu$, where $\Delta$ is the gap in the spectrum 
in the limit $\delta \mu =0$. When $\zeta> 1$, the entire
Fermi surface is gapped, \ie, when the thermal smearing of the Fermi
surface $\sim T$ is much smaller than the gap, no excitation can be
created out of the Fermi sphere. In this respect, the $\zeta >1$ case
resembles ordinary BCS superconductors. When $\zeta <
1$, particles can be excited from the gapless regions of the Fermi sphere.
We shall use below the interpolation formula proposed in Ref.~\cite{Jaikumar:2005hy},
which covers these asymptotic limits, \ie,
\bea 
\epsilon^{rg}_{\beta}(\zeta;T\le T_c) &=& 2f(\zeta) \epsilon_{\beta},
\nonumber\\
f(\zeta) &=& \frac{1}{\exp[(\zeta-1)\frac{\delta\mu}{T}-1]+1},
\eea
where the factor two comes from the two colors of quarks involved in the pairing.
Since the pairing pattern breaks the $SU(3)$ color symmetry, one of the
quark colors (say, blue) remains unpaired. Same-color pairing leads to
much smaller gaps in the spectrum of the quarks; these could be as
small as tens of keV, and therefore irrelevant in the neutrino-cooling
era where $T> 100$ keV in the core. The pairing in this case is
BCS-type, and we shall simply parameterize the modifications of the
Urca process on blue quarks as in the case of baryonic matter, \ie, 
\be
 \epsilon^{b}_{\beta}(T\le T_c) 
=\epsilon_{\beta}\exp\left(-\frac{\Delta_b}{T}\right).
\ee 
Note that $\epsilon^{\beta}_b = \epsilon^{\beta}_{rg}/2$ due to the
different number of colors involved in these cases. 

As in ordinary BCS superconductors, the pair correlations suppress the
specific heat of quarks (for $T \ll T_c$ exponentially); at $T=T_c$,
the specific heat jumps (increases) if the phase transition to the
superconducting phase is of second order. To model the modifications
of the specific heat, we adopt the fitted formula for the ratio of the
specific heat in the superconducting $c_S$ and normal $c_N$ phases
~\cite{M59} \bea\label{cv_blue} \frac{c_S}{c_N}(\tau) =
\left\{\begin{array}{cc} (12\pi/\gamma)
    (2\gamma\tau)^{-3/2}e^{-\pi/(\gamma\tau)}& 0\le \tau\le 0.3,
    \\-0.244+0.255\tau +2.431\tau^2 & 0.3< \tau\le
    1, \end{array}\right.  \eea where $\tau = T/T_c$ is the
temperature in units of the critical temperature and $\gamma =
1.781$. Equation~(\ref{cv_blue}) is applied to neutron, proton, and
blue-quark condensates. Note that Eq.~(\ref{cv_blue}) applies strictly
only to isotropic BCS condensates. The anisotropy of neutron
$P$-wave condensate somewhat modifies this ratio~\cite{Levenfish94}.

For the inhomogeneous red-green condensate, one needs to rescale the
critical temperature which changes with $\delta\mu$: \be
T_c(\zeta) \simeq T_{c0}\sqrt{1-\frac{4\mu}{3\Delta_0} \delta
  (\zeta)}, \ee where $\mu = (\mu_d+\mu_u)/2$, $\Delta_0 =
\Delta(\zeta=0)$, $T_{c0} = T_c(\zeta=0)$, and $\delta (\zeta) =
(n_d-n_u)/(n_d+n_u)$.  The fully gapped and gapless regimes behave
differently; in the presence of gapless modes, the specific heat has a
linear dependence on the temperature. A phenomenological way to model
this behavior is given by the relation \bea c_S^{rg}(\zeta;T\le T_c)
&=& f(\zeta) c_N^{rg}, \eea where $c_N^{rg}$ is the specific heat of
red-green unpaired quarks, taken as that for noninteracting quarks~\cite{Ipp:2003cj}
and $c_S^{rg}$ is the specific heat of pair-correlated quarks.  To
compute the specific heat of the electrons, we assume they form a
noninteracting, ultrarelativistic gas.

The hadronic matter in our models is composed of neutrons
(constituting $\sim 95\%$ of baryonic matter content), of protons
(with abundances of $\sim 5\%$ of baryonic matter) and electrons
(whose number is equal to that of protons by virtue of charge
neutrality). In the unpaired state, the dominant cooling agents for
such abundances are the modified Urca processes and neutral current
bremsstrahlung processes, included in standard fashion.  We use
emissivities derived for the free pion-exchange model of strong
interaction~\cite{FM79}, but reduce them by a constant factor 5 to
account for short-range repulsive component of the nuclear
force. (Note that medium modifications of pion dispersion can further
modify these rates and the resulting cooling substantially; see
Ref.~\cite{Schaab:1996gd}. However, no clear evidence of such
modification were found in experiments so far). The corresponding rates below
$T_c$ are exponentially suppressed, e.g., for the modified Urca
process we have 
\be \epsilon_{\beta\,{}\rm mod} (T\le T_c) =
\epsilon_{\beta\,{\rm mod}}
\exp\left(-\frac{\Delta_n(T)+\Delta_p(T)}{T}\right).  
\ee 
In addition, we include the process of electron bremsstrahlung on
nuclei in the crusts~\cite{Festa}. Following the latter
reference, we assume that the ions form a fluid throughout the crust 
(or the number of impurities is high) so that the emissivity scales as
$T^6$. If at some temperatures ions form a crystalline phase, the emissivities
scale roughly as $T^7$ (classical crystal) or as $T^8$ (quantum
crystal) and the bremsstrahlung emissivity is
parametrically suppressed~\cite{Kaminker}.

\begin{figure}[t] 
\begin{center}
\includegraphics[width=8.0cm,height=8.0cm]{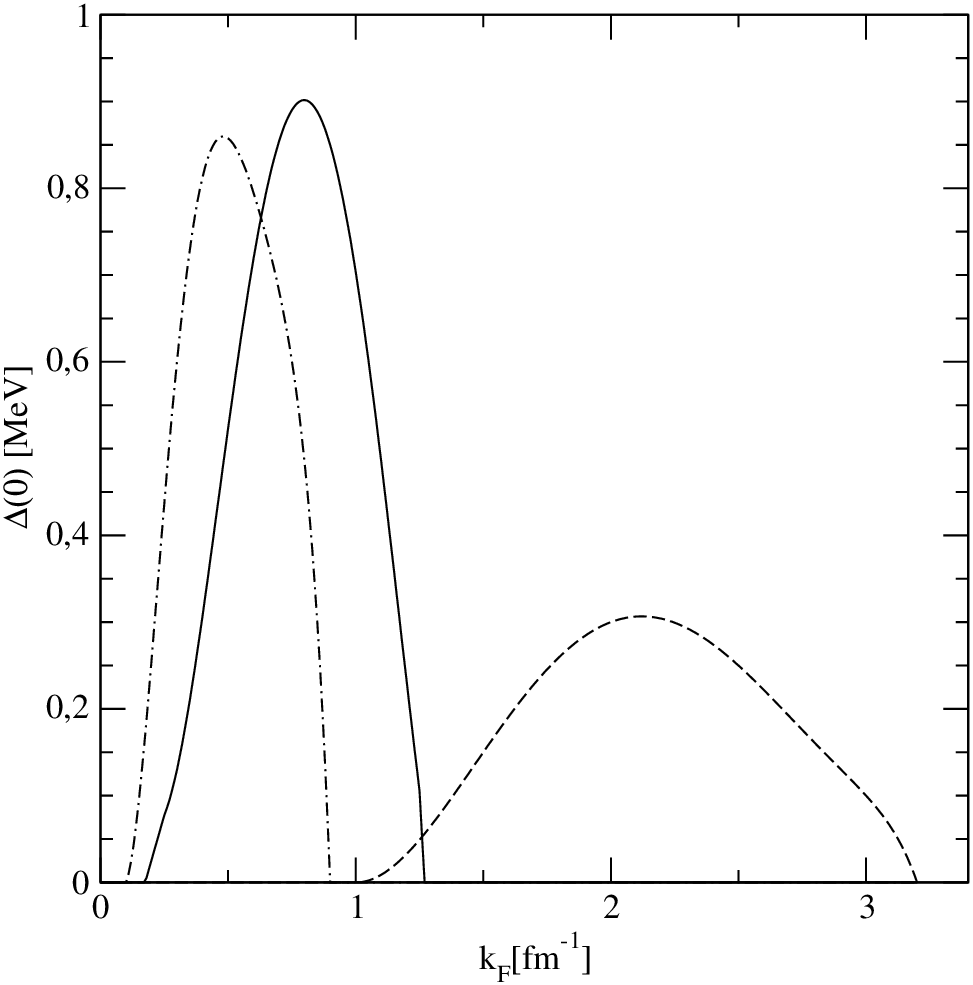}
\caption{ Dependence of the neutron  and proton pairing gaps on
  their Fermi momenta, (solid line  -  $^1S_0$ neutron
  gap~\cite{Wambach:1992ik},  dashed  line -   $^3P_2-^3F_2$ neutron 
gap~\cite{Baldo:1998ca}, and  dashed-dotted line -$^1S_0$ proton
pairing gap~\cite{Baldo:1992}).
}
\label{fig:2}
\end{center}
\end{figure}

In the superfluid/superconducting hadronic phases, the processes
of pair-breaking can contribute to the cooling ~\cite{Flowers:1976ux}. 
The neutral vector current processes are strongly suppressed by
multiloop processes~\cite{Leinson:2006gf,Sedrakian:2006ys,Kolomeitsev:2008mc,Steiner:2008qz},
whereas the axial-vector emission can be taken at one-loop level to a
good accuracy~\cite{Flowers:1976ux}.  Therefore, only the axial-vector
neutrino emission through the pair-breaking processes were included in the
simulations. The density-dependent zero-temperature pairing gaps
of neutrons and protons, that have been used in the simulations, are
shown in Fig.~\ref{fig:2}.
Their non-zero-temperature values can
be obtained in terms of the zero-temperature values
$\Delta(0)$ using the formula~\cite{M59}
\bea\label{gap_T} 
\frac{\Delta(\tau)}{\Delta(0)} = \left\{\begin{array}{cc}
  1- \sqrt{2 \gamma \tau} e^{-\pi/(\gamma\tau)}
& 0\le
   \tau\le 0.5, \\
\sqrt{3.016(1-\tau) -2.4(1-\tau)^2}
& 0.5< \tau\le  1 ,\end{array}\right.  \eea 
which reproduces the exact BCS result with an error 
of order of a percent.

\section{Thermal evolution}
\label{sec:3}

Our models were evolved in time, with the input described above, to
obtain the temperature evolution of the isothermal-interior.  The
isothermal-interior approximation is valid for timescales $t\ge 100$ yr,
which are required to dissolve temperature gradients by thermal
conduction.  Unless the initial temperature of the core is chosen too
low, the cooling tracks exit the nonisothermal phase and settle at a
temperature predicted by the balance of the dominant neutrino emission
and the specific heat of the core {\it at the exit temperature}. The
thermal conduction is effective enough to erase thermal gradients at
densities larger than  $10^{10}$ g cm$^{-3}$. The low-density envelope maintains
substantial temperature gradients throughout the entire evolution; the
temperature drops by about 2 orders of magnitude within this
envelope.  The isothermal-interior approximation relies further on the
fact that the details of the temperature gradients within the envelope
are unimportant if we are interested only in the surface
temperature. Models of the envelopes predict the scaling $T_s^4 = g_s
h(T)$, where $g_s$ is the surface gravity, and $h$ is some function
which depends on $T$, the opacity of crustal material, and its
equation of state.  We shall use the fitted formula $T_8 = 1.288
(T_{s6}^4/g_{s14})^{0.455}$~\cite{GPE:82}.  (Note that this relation
does not take into account either the potential strong magnetization
of neutron star envelopes or the presence of light elements in their
surface layers. The modifications due to these factors are discussed
in Refs.~\cite{Potekhin:2001id,Potekhin:1997mn}).

In the isothermal-interior approximation, the parabolic differential
equation for the temperature reduces to an ordinary differential
equation, 
\begin{figure*}[t] 
\begin{center}
\includegraphics[width=15.0cm,height=11.0cm]{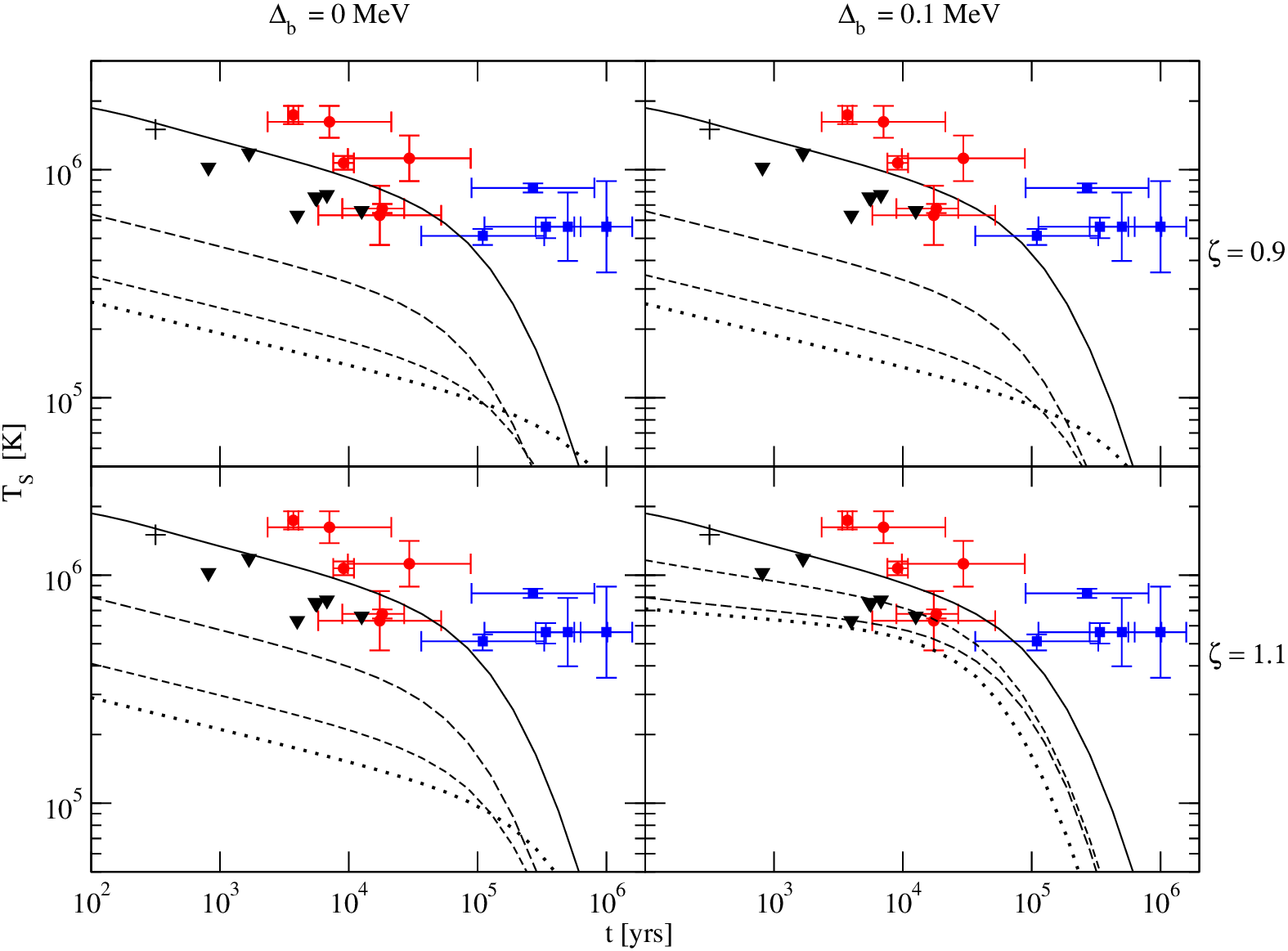}
\caption{ Time evolution of the surface temperature of four models
  with central densities (in units of $\rho_{14}$) 5.1 (solid line),
  10.8 (long-dashed line), 11.8 (short-dashed line), 21.0 (dotted
  line). The integral parameters of these models can be found in 
  Table~\ref{table:1}.  The observational data points in order of
  increasing age correspond to RX J0822-4247, 1E 1207.4-5209, RX
  J0002+6246, PSR 1706-44, PSR 0833-45 (Vela), PSR 0538+2817 (dots, red
  online), PSR 0656+14, PSR 1055-52, PSR 0633+1748 (Geminga), RX
  J1856.5-3754, RX J0720.4-3125 (squares, blue online).  The dots
  correspond to fits assuming H-atmospheres, squares -- to black-body
  fits. The data is taken from Ref.~\cite{Page:2009fu}. The triangles show
  in order of increasing age the current upper limits on the surface
  temperatures of PSR J0205+6449~\cite{Slane:2004jn} PSR
  J112439.1-591620~\cite{Hughes:2003yr}, supernova remnants (SNRs)
  G093.3+6.9, G315.4-2.3, G084.2+0.8, and
  G127.1+0.5~\cite{Kaplan:2004tm}, and RX
  J0007.0+73~\cite{Halpern:2004qg}. The pulsar in Cas A is shown by
  the plus sign~\cite{Heinke:2010cr,Shternin:2010qi}.  The upper two
  panels correspond to cooling when the red-green condensate has
  $\zeta = 0.9$, \ie, is not fully gapped; the lower panels correspond
  to $\zeta = 1.1$, \ie, the red-green condensate is fully gapped. The
  left two panels correspond to evolution with negligible blue-quark
  pairing ($\Delta_b =0$); the right two panels show the evolution for
  large blue pairing $\Delta_b=0.1$~MeV.  }
\label{fig:3}
\end{center}
\end{figure*}
\be \label{eq:master} C_V \frac{dT}{dt} = -L_{\nu} (T)-L_{\gamma}(T_s)
+ H (T), \ee where $L_{\nu}$ and $L_{\gamma}$ are the neutrino and
photon luminosities, $C_V$ is the specific heat of the core, and the
heating processes, which could be important in the photon cooling era,
are neglected, \ie, $H(T) = 0$ (see
Refs.~\cite{Schaab:1999as,Gonzalez:2010ta} for a summary of these
processes). The photon luminosity is given simply by the
Stephan-Boltzmann law $L_{\gamma} = 4\pi \sigma R^2T_s^4$, where
$\sigma$ is the Stephan-Boltzmann constant. The neutrino luminosities
$L_{\nu} (T)$ and the specific heat $C_V$ in Eq.~(\ref{eq:master}) are
spatial integrals of emissivities and local specific heats of
constituents over the volume of the isothermal-interior.  In
nonsuperfluid matter, the temperature dependence of the luminosities
and the specific heat in Eq. (\ref{eq:master}) can be factored out and
the remaining coefficients need to be computed only once for a given
stellar model. If matter is superfluid, the temperature {\it and} density
dependence of the pairing gaps (or critical temperatures) cannot be
factorized. Hence, we carry out the spatial integrations over the
volume of the isothermal interior at each time step to obtain the relevant
coefficients, which are now temperature dependent. Once the
coefficients are known, Eq. (\ref{eq:master}) can be advanced in time,
\eg, through the Runge-Kutta algorithm.
\begin{figure*}[t] 
\begin{center}
\includegraphics[width=15.0cm,height=11.0cm]{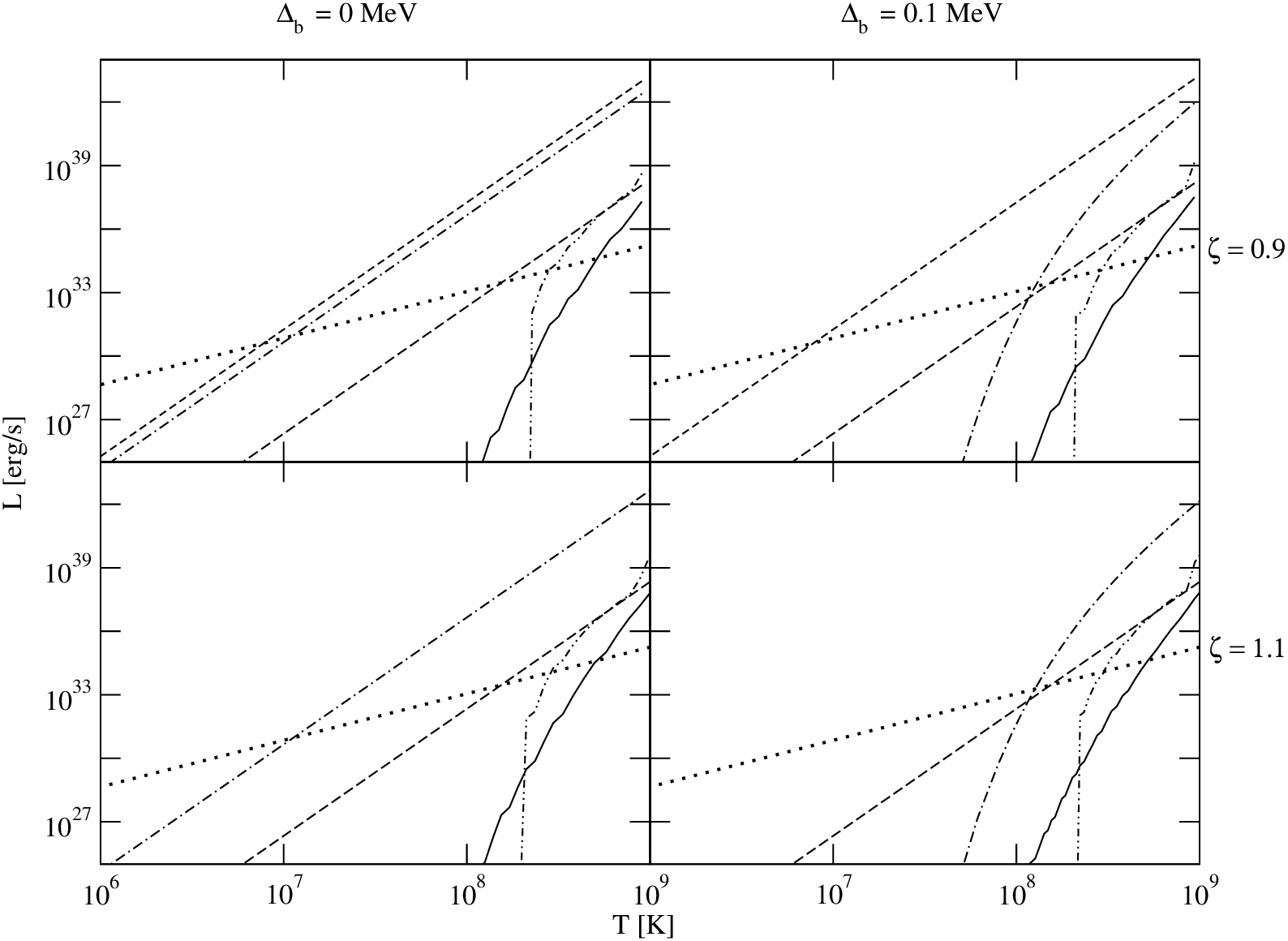}
\caption{ Dependence of total luminosities of contributing processes 
on the core temperature: solid lines - Urca and bremsstrahlung
processes in hadronic phase(s), long-dashed line - electron
bremsstrahlung in the crusts, short dashed  and dashed-dotted lines 
-  Urca process with red-green and blue quarks, respectively,
dashed-double-dotted lines - pair-breaking processes in the hadronic
phases, dotted lines - photon emission from the surface. 
The underlying model corresponds to the 1.93 solar-mass model 
in Table~\ref{table:1}.
}
\label{fig:4}
\end{center}
\end{figure*}
The results of integration of Eq.~(\ref{eq:master}) are shown in
Fig.~\ref{fig:3}, where we display the dependence of the (redshifted)
surface temperature on time. Each panel of Fig.~\ref{fig:3} contains
cooling tracks for the same set of four models introduced in
Sec.~\ref{sec:3}; the cooling tracks for the purely hadronic model
(solid lines) are the same in all four panels. The panels differ in
the values of microphysics parameters, which characterize the pairing
pattern in quark matter.  Specifically, the two panels in the left
column correspond to the case where the blue-quark pairing is
negligible (\ie, the pairing is on a scale much smaller than the
smallest energy scale involved, typically the core temperature). The
two panels in the right column correspond to the case where the gap
for blue quarks is large, $\Delta_b = 0.1$ MeV. The panels in the
upper and lower rows are distinguished by the value of the $\zeta$
parameter.  [We use the values $\zeta = 0.9$ (upper row) and $\zeta =
1.1$ (lower row)].

Figure~\ref{fig:4} shows the dependence of the neutrino luminosities
of contributing processes on the core temperature for the star model 
with central density $\rho_{c,14} = 11.8$.  It is complementary to
Fig.~\ref{fig:3} in the sense that it allows us to identify the
dominant cooling agent for each temperature and, hence, via
Fig.~\ref{fig:3}, for each epoch.  Clearly, the cooling
mechanism with largest luminosity determines the rate of decrease of 
the temperature. Since no heating processes are included in our study, 
the asymptotic rate of the cooling is determined by the photon
luminosity.  The dominant cooling agent in the neutrino emission era 
depends on the microphysics input as explained in caption of Fig~\ref{fig:4}.
\begin{figure}[t] 
\begin{center}
\includegraphics[width=7.0cm,height=7.0cm]{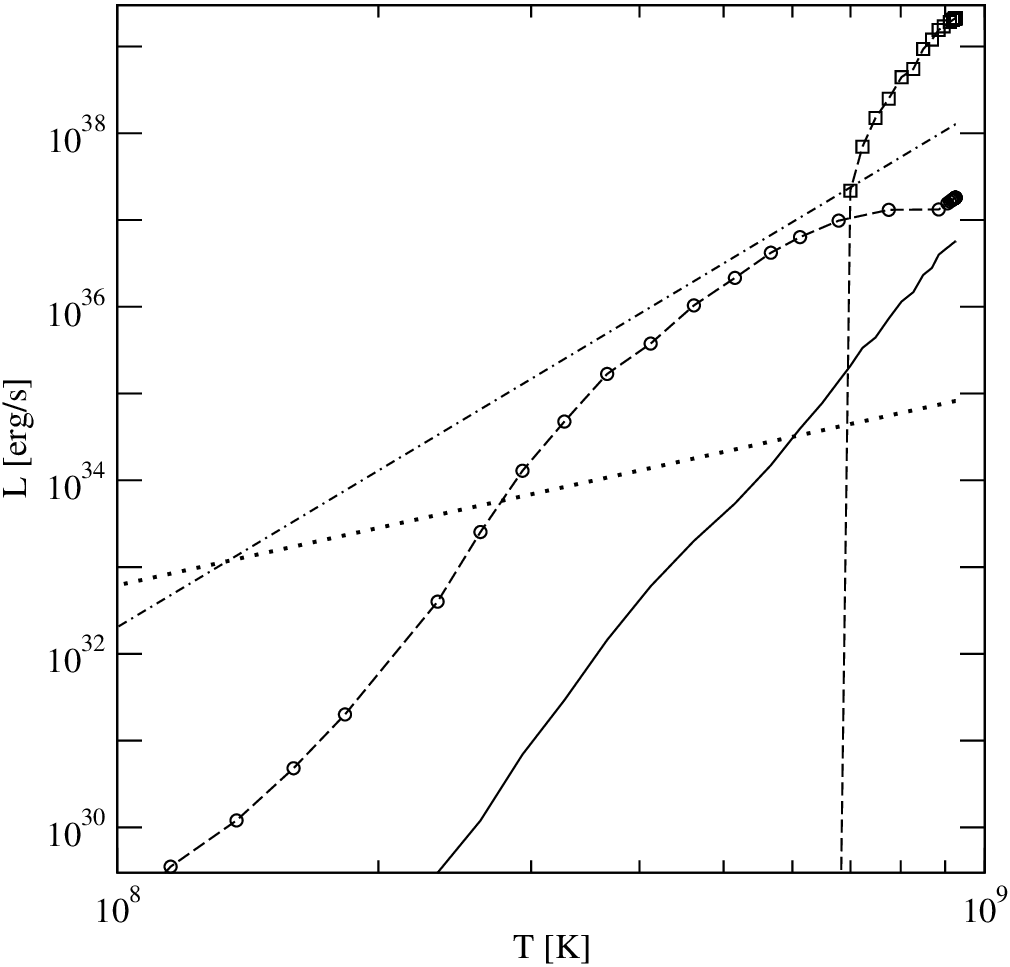}
\caption{ Dependence of total luminosities on the core temperature 
for a 1.1 $M_{\odot}$ hadronic star.  Solid line - the
luminosity of two-nucleon processes (modified Urca and bremsstrahlung), dashed-dotted line - the
bremsstrahlung luminosity in the crust,  dotted line - the photon
luminosity of the surface, circles - neutron pair-breaking process in the
crust, squares - neutron and proton pair-breaking processes in the core.
}
\label{fig:5}
\end{center}
\end{figure}

A general trend seen in each panel of Fig.~\ref{fig:3} is that the
heavier the star, the faster is its cooling by neutrinos. As a
consequence, the heavy stars enter the photon cooling stage later than
the lighter ones. In the late photon cooling era, the arrangement of
the cooling curves is inverted, \ie, the heavier stars are ``hotter''
than the lighter stars. (The photon luminosity of a star depends on
its radius. Therefore, the heavier stars, which are more compact than
the light ones, have smaller photon luminosities than the light stars
of the same surface temperature; however, the cooling dynamics
dominates over this static effect.) The fast cooling of massive stars
in the neutrino emission era is driven by the quark Urca process (see
Fig.~\ref{fig:4}). Since the neutrino luminosity in
Eq.~(\ref{eq:master}) is an integral quantity, it scales with the size
of the quark core $L_{\nu}\propto [R^{q}]^3$, and therefore increases 
with the size of the quark core. The scatter of cooling
curves in the $T-t$ plane suggests that the experimental data could be
naturally accommodated in the models that have quark cores whose size
changes as the central density of the model (and the net mass) are
changed. In such an approach, the dichotomy in cooling behavior of
neutron stars is attributed to the different masses of these objects:
the lighter ones not containing quark matter cool more slowly than the
heavier stars having quark cores. Note, however, that other factors, e.g., surface
magnetic fields or the envelope structure, could be responsible for
differences observed in the surface temperatures of isolated neutron
stars as well.

Superimposed on this general trend are the variations that arise from 
different functional dependence of the Urca emissivity of quarks on
the temperature under different assumptions about the pairing pattern of
quarks. We explore the variation of the parameters $\Delta_b$ and
$\zeta$.  The fastest neutrino cooling occurs for $\Delta_b = 0$ 
and $\zeta =0.9$ (the upper left panels in Figs.~\ref{fig:3} 
and \ref{fig:4}). Indeed, in this case the
Urca process is unsuppressed for all colors of quarks; the 
red-green quarks radiate neutrinos at rates comparable to 
unpaired quark matter, since there exist gapless excitations near the
Fermi surfaces for $\zeta < 1$.  The next-fastest cooling scenario
corresponds to the choice $\Delta_b = 0.1$ MeV and $\zeta = 0.9$ 
(the upper right panel in Figs.~\ref{fig:3} and Fig.~\ref{fig:4});  
here, the contribution of blue quarks is 
suppressed for $T< 10^9$ K, whereas the radiation
from red-green quarks is unsuppressed as before.
Therefore, the
neutrino luminosity is reduced by about 1/3 from the case where 
all quark colors radiate at the rate of unpaired quark matter. The effect
of a large blue-quark gap is clearly seen in Fig.~\ref{fig:4} where the
initially power-law drop of the luminosity with temperature changes to
exponential. Let us turn to the case where red-green quarks are gapped
over the entire Fermi surface, \ie,  $\zeta> 1$. Because their
gap is substantially larger than the blue-quark gap, their
neutrino luminosity  is suppressed at a much larger temperature. 
(This corresponds to the case $\zeta = 1.1$, which is illustrated in  
the two panels in the lower row in Figs.~\ref{fig:3} and \ref{fig:4}.) 
In the case where $\Delta_b=0$
(lower left panel), we find the third-fastest neutrino-cooling
scenario, in which
 the neutrino emission is dominated by the blue quarks and is 
about $30\%$ of the  rate that would be emitted by quarks of all
colors. Finally, the slowest cooling scenario emerges when
$\Delta_b=0.1$ MeV
and $\zeta = 1.1$ (lower right panel). The neutrino emission of quarks
of all colors is shut off below the critical temperature of blue
quarks. The photon cooling from the surface and the bremsstrahlung in
the crust are dominant close to this temperature, but the photon
luminosity takes over as the temperature drops slightly further (see
Fig.~\ref{fig:4}, lower left panel).  

Figure \ref{fig:4} offers some further insight into the relative
importance of various cooling processes.  Specifically, the neutrino
emission from the hadronic core with $L_{core}\propto T^8$ in the
unpaired matter is an unimportant source of cooling except in the slow
cooling model, because of the exponential suppression due to the gaps
in neutron and proton quasiparticle spectra. The bremsstrahlung
process in the crusts with $L_{crust}\propto T^6$ is numerically
unimportant except for a limited range of temperatures prior to the
transition to the photon cooling era. The fact that over long periods
of time a single cooling process is dominant allows one to deduce the
slope of the cooling curves, provided the specific heat capacity of
the fermions scales linearly with temperature $c_V = aT$ and the
nontrivial temperature dependence arising from pairing effects can be
neglected. Writing the luminosity of a process as $L = bT^{n+2}$,
Eq. (\ref{eq:master}) can be integrated to obtain \be t =t_0+
\frac{a}{bn}T(t) ^{-n} \left(1-\frac{T(t)^{n}}{T_0 ^{n}}\right), \ee
where $t_0$ is the initial time and $T_0 = T(t_0)$ is the initial
temperature. The expression in the braces is unity for $T(t) \ll T_0$,
so one finds the scaling $t\sim T^{-n}$. Thus, the neutrino-cooling
time via quark Urca process scales as $T^{-4}$, while the cooling time
via photons scales as $T^{-1/5}$ for the surface-core temperature
relation adopted above.

In is further instructive to examine the relative importance of
different hadronic processes in a purely hadronic $1.1 M_{\odot}$ mass
star. As seen in Fig.~\ref{fig:5}, in the temperature domain $7\times
10^8\le T\le T_c$ the neutrinos are produced predominantly by the
pair-breaking processes in the core. At lower temperatures the crust
electron bremsstrahlung and, to some extent, the pair-breaking
bremsstrahlung in the crustal superfluid are the dominant
processes. At about $T\sim 10^8$ K the neutrino emission era ends and
the photon cooling from the surface of the star becomes the dominant
cooling process.

A comparison with the measured temperatures of neutron stars suggests
that slow cooling models ($\Delta_b = 0.1$ and $\zeta = 1.1$) describe
young, cool objects. The temperatures of the fast cooling models are
well below the current measurements and upper limits. Our 1.1
$M_{\odot}$ hadronic model fails to account for the data on old and
hot stars. There are several physical factors that may influence this
outcome: (i)~the rates of the neutrino bremsstrahlung of electrons in
the crust may be overestimated due to the our assumption that the ions
form a fluid~\cite{Kaminker}; (ii)~if the crust bremsstrahlung is
subdominant due to crystalline nature of the lattice, the
pair-breaking process are the most important cooling agents in a wider
temperature range and these depend sensitively on the values of
not-well-constrained pairing gaps in the baryonic matter;
(iii)~variations in the magnetic fields and the surface composition of
the neutron stars.

\section{Conclusions}
\label{sec:4}
Motivated partly by the recent observation of a massive compact object
in a binary system~\cite{Demorset:2010} and partly by the recent
development of sequences of stable massive hybrid stars with realistic
input equations of state~\cite{Ippolito:2007hn,Knippel:2009st}, we
have modeled the thermal evolution of compact stars containing quark
cores. These sequences of stable stars allow for a transition from
hadronic to quark matter in massive stars ($M> 1.85 M_{\odot}$) with
the maximal mass of the sequence $\sim 2 M_{\odot}$.

Our study has concentrated on the changes in the cooling
behavior of the compact objects as the central density (and therefore
macro-parameters such as the mass) is varied in a wide range
along a sequence which has low-mass ($M\le 1.85 M_{\odot}$) purely
hadronic members and high-mass  ($M> 1.85 M_{\odot}$) stars containing
quark cores. We have seen that different cooling
scenarios arise depending on the assumption of the magnitude
of the pairing gaps in quark matter, including the possibility of
emergence of a gapless phase in $u$-$d$ condensate of quarks.

We find that (i) the neutrino-cooling is slow for hadronic stars and 
becomes increasingly fast with an  increase of the size of the quark
core, in those scenarios where there are unpaired quarks or
gapless excitations in the superconducting quark phase. The
temperature scatter
of the cooling curves in the neutrino cooling era is significant and can
explain the observed  variations in the surface temperature data of 
same age neutron stars. (ii) If quarks of all colors have gapped Fermi
surfaces, the neutrino cooling shuts off early, below the pairing
temperature of blue quarks; in this case, the temperature 
spread of the cooling
curves is not as significant as in the fast cooling scenarios. (iii) 
As the stars evolve into the photon cooling stage the temperature
distribution is inverted, \ie, those stars that were cooler in the 
neutrino-cooling era are hotter during the photon cooling stage.

Before closing, we briefly state some potential modifications to the
results above and some further refinements of the models presented
above. In this work, we concentrated on the phase with two-flavor
pairing. If the in-medium masses of strange quarks are low, they could
appear at low densities relevant to compact stars; the cooling of
objects containing condensates with strange quarks has been studied
elsewhere~\cite{Alford:2004zr}. Since the fermionic excitations in
these denser phases differ from those in our model, their cooling
significantly differs from the  cooling behavior we observe.

The late-time evolution of compact objects will be affected by the
heating in their interiors, as indicated in Eq.~(\ref{eq:master}). 
Heating processes change the thermal history in the photon cooling, 
implying a less steeper than  $\propto T^{-1/5}$ slope of temperature 
decay with time~\cite{Schaab:1999as,Gonzalez:2010ta}.
 While important, these processes are unlikely to
affect the present results and conclusion, which are mainly concerned with
the neutrino emission era.

Although our models have realistic density profiles derived from
microscopic equations of state, we have assumed constant pairing gaps
throughout the color superconducting phases. This is not the case,
since the density of states at the Fermi surface and running coupling
of strong interactions  (both of which control the strength of pairing
interactions) vary with density. As the density increases, the density
of states increases as 1/3 power of density, whereas the strong
coupling constant decreases logarithmically.  These variations would
be of minor importance unless the $\zeta$ function occurs to cross
$\zeta =1$ boundary separating the gapless and gapped phases. If this
occurs, the core shell where $\zeta <1$ will be the predominant
neutrino emitter; physically, instead of the entire core volume the
effective radiation volume will amount to that of the shell where
$\zeta < 1$. For any given model, this will reduce the neutrino
luminosity of red-green quarks. A density-dependent $\zeta$ parameter,
combined with other parameters discussed above, can be used to
fine-tune the models to fit the observational data. In particular, our
models can be applied to describe the recently observed cooling behavior
of the neutron star in Cassiopeia
A~\cite{Heinke:2010cr,Shternin:2010qi}.

\acknowledgements
We thank M.~Alford, D.~B.~Blaschke,  J. W. Clark, X.-G.~Huang, B.~Knippel,  
L.~ Rezzolla, D.~H.~Rischke, A.~ Schmitt, K.~Schwenzer,
and  I.~Shovkovy for useful discussions and B.~Knippel for 
collaboration at the initial stages of this project.

\end{document}